\documentclass[]{aastex631}
\usepackage{amsmath}
\usepackage{multirow}
\usepackage{hyperref}
\usepackage{CJK}

\shorttitle{Multi-Spectral Sirens}
\shortauthors{Li et al.}

\graphicspath{{./}{figures/}}

\begin{document}

\begin{CJK*}{UTF8}{gbsn}

\title{Multi-spectral Sirens: Gravitational-wave Cosmology with (Multi-) subpopulations of Binary Black Holes}

\author[0000-0001-5087-9613]{Yin-Jie Li （李银杰）}
\affiliation{Key Laboratory of Dark Matter and Space Astronomy, Purple Mountain Observatory, Chinese Academy of Sciences, Nanjing 210023, People's Republic of China}

\author[0000-0001-9120-7733]{Shao-Peng Tang（唐少鹏）}
\affiliation{Key Laboratory of Dark Matter and Space Astronomy, Purple Mountain Observatory, Chinese Academy of Sciences, Nanjing 210023, People's Republic of China}

\author[0000-0001-9626-9319]{Yuan-Zhu Wang（王远瞩）}
\affiliation{Institute for Theoretical Physics and Cosmology, Zhejiang University of Technology, Hangzhou, 310032, People's Republic of China}
\affiliation{Key Laboratory of Dark Matter and Space Astronomy, Purple Mountain Observatory, Chinese Academy of Sciences, Nanjing 210023, People's Republic of China}

\author[0000-0002-8966-6911]{Yi-Zhong Fan（范一中）}
\affiliation{Key Laboratory of Dark Matter and Space Astronomy, Purple Mountain Observatory, Chinese Academy of Sciences, Nanjing 210023, People's Republic of China}
\affiliation{School of Astronomy and Space Science, University of Science and Technology of China, Hefei, Anhui 230026, People's Republic of China}
\email{The corresponding author: yzfan@pmo.ac.cn (Y.Z.F)}

\begin{abstract}
The cosmic expansion rate can be directly measured with gravitational-wave (GW) data of the compact binary mergers by jointly constraining the mass function of the population and the cosmological model via the so-called spectral sirens.  Such a method relies on the features in the mass functions, which may originate from some individual subpopulations, and hence become blurred/indistinct due to the superposition of different subpopulations. In this work we propose a novel approach to constrain the cosmic expansion rate with subpopulations of GW events, named multi-spectral sirens. The advantage of the multi-spectral sirens compared to the traditional spectral sirens is demonstrated by the simulation with the mock data. The application of this approach to the GWTC-3 data yields $H_0=73.3^{+29.9}_{-25.6}~{\rm Mpc}^{-1}~{\rm km}~{\rm s}^{-1}$ (median and symmetric 68.3\% credible interval), which is about 19\% tighter than the result inferred with the traditional spectral sirens utilizing a PowerLaw+Peak mass function. The incorporation of the bright standard siren GW170817 with a uniform prior in [10,200] (log-uniform prior in [20,140]) ${\rm Mpc}^{-1}~{\rm km}~{\rm s}^{-1}$ gives $H_0=71.1^{+15.0}_{-7.5}~(70.3^{+12.9}_{-7.1})~{\rm Mpc}^{-1}~{\rm km}~{\rm s}^{-1}$ (68.3\% confidence level), corresponding to an improvement of $\sim26\%$ (23\%) with respect to the measurement from sole GW170817.
\end{abstract}


\section{Introduction} \label{sec:intro}

The discrepancy in the values of the Hubble constant $H_0$ as determined by early and late Universe observations \citep{2016A&A...594A..13P, 2019ApJ...876...85R, 2019NatAs...3..891V} has attracted wide attention over the past decade and remains unresolved. Gravitational waves (GWs) from coalescing compact binaries (CBCs) provide measurements of the luminosity distances of sources. As long as the redshift of a nearby source is known, $H_0$ can be robustly inferred \citep{1986Natur.323..310S, 2005ApJ...629...15H}. Such a goal has been firstly achieved with the multi-messenger data of a binary neutron star (BNS) merger event GW170817/GRB 170817A/AT2017gfo \citep{2017PhRvL.119p1101A,2017ApJ...848L..12A,2017Natur.551...85A}.
In the absence of the redshift measurements, GWs from CBCs alone can also constrain cosmological parameters through the use of so-called spectral sirens \citep{2012PhRvD..85b3535T, 2019ApJ...883L..42F, 2021ApJ...908..215Y, 2021PhRvD.104f2009M, 2022PhRvL.129f1102E, 2023ApJ...949...76A, 2024arXiv240402522M, 2024arXiv240402210F}. In this scenario, distinct and recognizable features in the CBC mass functions are helpful for better inferring $H_0$. These features include, e.g.,  the lower-mass gap between the heaviest NSs and the lightest black holes (BHs) \citep{1974PhRvL..32..324R, 1996ApJ...470L..61K, 1998ApJ...499..367B, 2010ApJ...725.1918O, 2011ApJ...741..103F, 2021ApJ...923...97L, 2022ApJ...931..108F,2024PhRvD.109d3052F}, as well as the higher-mass gap resulting from (pulsational) pair-instability supernova (PPISN) explosions \citep{2017ApJ...836..244W, 2021ApJ...912L..31W}, 
i.e., the pair-instability mass gap (PIMG).

However, these features in the mass functions may be indistinct due to the superposition of subpopulations from different formation channels \citep{2021ApJ...910..152Z, 2022ApJ...933L..14L, 2022ApJ...928..155T, 2023ApJ...955..127C, 2023arXiv230401288G}. 
For instance, the PIMG could be populated by hierarchical mergers \citep{2021ApJ...915L..35K, 2022ApJ...941L..39W, 2024PhRvL.133e1401L, 2024arXiv240601679P}, which may blur its lower edge and make it indistinguishable \citep{2021ApJ...913...42W, 2021ApJ...913L...7A, 2023PhRvX..13a1048A}. Hence, its application in the spectral sirens study is hampered.

In this work, we introduce a multi-spectral sirens method which decomposes the mass spectra of subpopulations from the overall population of BBHs/BHs, to make the features in the mass function more distinct and hence potentially improve the accuracy of the cosmological parameter measurements.
subpopulations of BBHs/BHs can be identified through the analysis of additional parameters such as spins and redshifts \citep{2021ApJ...910..152Z,2021NatAs...5..749G}. This allows for clearer identification of features in the mass functions of these subpopulations, which in turn improves the capability of spectral sirens in measuring cosmological parameters.
Our previous investigations \citep{2024PhRvL.133e1401L} have successfully identified two distinct subpopulations of BHs within the GWTC-3 dataset \citep{2019PhRvX...9c1040A, 2021PhRvX..11b1053A, 2021arXiv210801045T, 2021arXiv211103606T}, employing semi-parametric models \citep[see also][for astrophysically motivated analysis]{2022ApJ...941L..39W}. The first subpopulation displays a sharp cut-off at $\sim 40 M_{\odot}$, consistent with the PIMG \citep{2019ApJ...887...53F}. The second subpopulation features a distinct spin-magnitude distribution (peaking at approximately 0.7), suggesting origins from hierarchical mergers \citep{2017ApJ...840L..24F, 2017PhRvD..95l4046G}. These findings have been supported by subsequent studies \citep[e.g.,][]{2024arXiv240409668L, 2024arXiv240601679P, 2024arXiv240603257G}.

The remainder of the article is organized as follows: Section~\ref{sec:method} introduces the analysis framework and the novel population model employed for multi-spectral sirens. Section~\ref{sec:sim} discusses the enhancements of multi-spectral sirens over traditional spectral sirens. In Section~\ref{sec:real}, we present the results constrained by the GWTC-3 data. Discussions are provided in Section~\ref{sec:dis}.

\section{Methods}\label{sec:method}

\subsection{Population models}
Flexible mixture models are well suited for modeling subpopulations of BHs \citep{2024PhRvL.133e1401L, 2023arXiv230401288G, 2024arXiv240409668L}. In this study, we use a basic mixture model which fits the subpopulations within the component-mass versus spin-magnitude distribution \citep{2024PhRvL.133e1401L}. When the data get significantly enriched, one can use more sophisticated mixture models, e.g., models that incorporate the mass versus spin-orientation distribution (refer to the Supplementary of \citet{2024PhRvL.133e1401L} and also see \citet{2023arXiv230401288G} and \citet{2024arXiv240409668L}) and the mass versus spin versus redshift distribution \citep{2024arXiv240603257G}.

Employing a mixture model with an adequate number of components is beneficial when dealing with distinct subpopulations, as represented by 
\begin{equation}\label{eq:mix}
\pi(m,\chi|\mathbf{\Lambda})=\sum_{i=1}^{n}{\pi_i(m|\mathbf{\Lambda})\pi_i(\chi|\mathbf{\Lambda})r_i},
\end{equation}
where $m$ and $\chi$ denote the source-frame component mass and spin magnitude, respectively, while $r_i$ represents the mixture fraction of the $i$-th subpopulation. 
According to \citet{2024PhRvL.133e1401L}, for the currently available data, a model with two components ($n=2$) adequately captures the distribution of component mass versus spin magnitude.

In light of the data from GWTC-3, we adopt a non-parametric approach to model the component-mass distribution, which mitigates the systematic uncertainty that would otherwise arise from mis-modeling of parametric model. For this purpose, we utilize the \textsc{PowerLawSpline} model \citep{2022ApJ...924..101E}, since it has a concise formula, which reads
\begin{equation}
\pi_i(m | {\bf \Lambda}_i)=\mathcal{PS}(m|\alpha_i,m_{{\rm min},i},m_{{\rm max},i},\delta_{{\rm low},i},\delta_{{\rm up}, i},f_i(m;\{f_i^j\}_{j=0}^{N_{\rm knot}})).
\end{equation}
Here, $m_{{\rm min},i}$ and $m_{{\rm max},i}$ are the minimum and maximum mass cut-offs, $\alpha_i$ is slope index, $\delta_{{\rm low},i}$ and $\delta_{{\rm up}, i}$ are the smooth scale at lower and upper edges as defined in the smoothing function,
\begin{equation}\label{eq:sms}
    S(m|m_{\rm min},\delta_{\rm low},m_{\rm max},\delta_{\rm up})=
    \left\{
	\begin{aligned}
	&0 &\quad (m<m_{\rm min})\\
    &[f(m-m_{\rm min},\delta_{\rm low})+1]^{-1}   & \quad (m_{\rm min}<m<m_{\rm min}+\delta_{\rm low})\\
    &1 & \quad (m_{\rm min}+\delta_{\rm low}<m<m_{\rm max}-\delta_{\rm up})\\
    &[f(m_{\rm max}-m,\delta_{\rm up})+1]^{-1}   & \quad (m_{\rm max}-\delta_{\rm up}<m<m_{\rm max})\\
	&0 &\quad (m_{\rm max}<m).
 \end{aligned}
	\right	.
\end{equation}
where $f(x,\delta_{\rm m})={\rm exp}({\delta_{\rm m}}/{x}+{\delta_{\rm m}}/{(x-\delta_{\rm m})})$.
To streamline the model, we set the smooth scales to 0 for the second subpopulation. This choice is motivated by the limited size of the events contributing to this subpopulation, as indicated by the currently available data \citep[see][]{2024PhRvL.133e1401L}. Consequently, the impact of the smooth function on the second subpopulation would be negligible. A smooth function may be more appropriate for this subpopulation when the sample has been significantly extended.
Following \cite{2022ApJ...924..101E}, we use 15 knots located linearly in the logarithm space within [5, 100] $M_{\odot}$ for the perturbation function $f_i$, and restrict the perturbation to zero at the minimum and maximum knots. The spin-magnitude distribution for $i$-th component is $\pi_i(\chi | {\bf \Lambda}_i)=\mathcal{G}(\chi| \mu_{\chi,i}, \sigma_{\chi,i}, 0, 1)$, a truncated Gaussian with peak $\mu_{\chi,i}$ and width $\sigma_{\chi,i}$ bounded in [0,1]. Then the overall population model reads
\begin{equation}\label{eq:pop}
\pi(\lambda|{\bf \Lambda})=A({\bf \Lambda})p(m_2/m_1|{\bf \Lambda})\pi(m_1,\chi_1|{\bf \Lambda})\pi(m_2,\chi_2|{\bf \Lambda})\pi(\cos\theta_1,\cos\theta_2|{\bf \Lambda}),
\end{equation}
where {\bf $\lambda=(m_1,m_2,\chi_1,\chi_2,\cos\theta_1,\cos\theta_2)$}, $A({\bf \Lambda})$ is the normalization factor, and $p(m_2/m_1|{\bf \Lambda})$ is the pairing function. Note that the pairing function may also be total-mass dependent \citep{2016ApJ...824L..12O}, however, it is vastly degenerated with the component-mass function \citep{2020ApJ...891L..27F}. Therefore, we only account for the mass-ratio dependent pairing function, since the ignorance of the total-mass dependent pairing have little effect on the inferred mass distribution for current data \citep{2022ApJ...941L..39W}. The spin-orientation distribution is the same as \textsc{DefaultSpin} model of \cite{2023PhRvX..13a1048A}, i.e.,
\begin{equation}\label{eq_tilt}
 \pi(\cos\theta_1,\cos\theta_2 |\zeta,\sigma_{\rm t})=\mathcal{U}(\cos\theta_1,\cos\theta_2 |-1,1)(1-\zeta)+\mathcal{G}(\cos\theta_1,\cos\theta_2 |1,\sigma_{\rm t},-1,1)\zeta,
\end{equation}
where $\mathcal{U}$ is a uniform distribution within (-1,1), and $\zeta$ is the mixture fraction of the nearly-aligned assembly. 

Following \cite{2023ApJ...949...76A},
the merger rate density as a function of redshift reads \citep{2014ARA&A..52..415M}, 
\begin{equation}\label{psi_z}
R(z|\gamma,\kappa,z_{\rm p})=[(1+z_{\rm p})^{(\gamma+\kappa)}+1](1+z)^{\gamma}/[(1+z)^{(\gamma+\kappa)}+(1+z_{\rm p})^{(\gamma+\kappa)}].
\end{equation}
It is characterized by a low-redshift power-law slope $\gamma$, a peak at redshift $z_{\rm p}$, and a high-redshift power-law slope $\kappa$ after the peak. Therefore the redshift distribution of BBHs is,
\begin{equation}\label{p_z}
\pi(z|\gamma,\kappa,z_{\rm p},H_0,\Omega_{\rm m}) = \Phi_{\rm c}\frac{\mathrm{d}V_{\rm c}(H_0,\Omega_{\rm m})}{(1+z)\mathrm{d}z}R(z|\gamma,\kappa,z_{\rm p}),
\end{equation}
where $V_{\rm c}$ is the comoving volume, and $\Phi_{\rm c}$ is the normalization constant.
All the descriptions of the hyper-parameters and the priors are summarized in Table~\ref{prior}.

\subsection{Cosmological model}
We use a flat $\Lambda$CDM cosmological model in this work, and assume the dark energy density is constant during the cosmic expansion. Then the function of luminosity distance $D_L$ and redshift $z$ is represented as \citep{2023ApJ...949...76A}
\begin{equation}
D_L(z)=\frac{c(1+z)}{H_0}\int_{0}^{z}{[\Omega_{\rm m}(1+x)^3+1-\Omega_{\rm m}]^{-1/2} \mathrm{d}x}={\rm F}(z|H_0,\Omega_{\rm m}),
\end{equation}
where $\Omega_{\rm m}$ is the present-day dimensionless matter density, and $H_0$ is the Hubble constant.
The GW signal enables the measurement of detector-frame masses of BBHs and the luminosity distance (i.e., $M_1$, $M_2$, $D_{L}$). Subsequently, the cosmology (with parameters  $H_0$ and $\Omega_{\rm m}$) allow for the calculation of source-frame masses. These are determined by the formula $m_{1,2} = M_{1,2} / (1 + z(D_L)) = M_{1,2} / (1 + {\rm F}^{-1}(D_L|H_0, \Omega_{\rm m}))$. 

\subsection{Hierarchical inference framework}
We use hierarchical Bayesian inference to jointly fit the source population and cosmological models.
For the given data $\{d\}$ from $N_{\rm det}$ GW detections, the likelihood \citep{2023PhRvX..13a1048A} for the hyperparameters $\boldsymbol{\Lambda}$ is
\begin{equation}\label{eq_llh}
\mathcal{L}(\{d\} |\boldsymbol{\Lambda})\propto N^{N_{\rm det}}e^{-N{\xi(\boldsymbol{\Lambda})}}\prod_{i=1}^{N_{\rm det}}\int{\mathcal{L}(d_i|\theta_i)\pi(\theta_i|\boldsymbol{\Lambda})d\theta_i},
\end{equation} 
where $N$ is the number of mergers over the surveyed space-time volume, and $\xi(\boldsymbol{\Lambda})$ means the detection fraction. 
The methodology for computing $\mathcal{L}(d_i|\theta_i)$ and $\xi(\boldsymbol{\Lambda})$ is detailed in \citet{2021ApJ...913L...7A}. It should be noted that we do not incorporate the spin-dependent selection bias, which is considered to have a negligible effect on our results (see \citet{2021ApJ...921L..15G} and \citet{2024PhRvL.133e1401L} for comprehensive discussions). For the estimation of the posterior distributions of the hyper-parameters, we employ the \textit{Pymultinest} sampler \citep{2016ascl.soft06005B}.

\begin{table}[htpb]
\begin{ruledtabular}
\caption{Hyper-parameters and Priors for the population and cosmological models}\label{prior}
\begin{tabular}{lccc}
\multirow{2}{*}{Descriptions}   & \multirow{2}{*}{parameters}  & \multicolumn{2}{c}{priors}  \\
\cline{3-4}
&&1st component & 2nd component \\
\cline{1-4}
slope index of the mass function  & $\alpha_i$ & U(-4,8) & U(-4,8) \\
smooth scale of the $i$-th mass lower edge& $\delta_{{\rm low},i}[M_{\odot}]$ &U(0,10)&0\\
smooth scale of the $i$-th mass upper edge& $\delta_{{\rm up},i}[M_{\odot}]$ &U(0,10)&0\\
 minimum mass of the $i$-th mass function&$m_{{\rm min},i}[M_{\odot}]$ & U(2,60)&U(2,60)\\
 maximum mass of the $i$-th mass function&$m_{{\rm max},i}[M_{\odot}]$ & U(20,100) &U(20,100)\\
interpolation values of perturbation for $i$-th mass function &$\{f_i^j\}_{j=2}^{14}$ &$\mathcal{N}(0,1)$ &$\mathcal{N}(0,1)$ \\
mass constraints & & \multicolumn{2}{c}{$m_{{\rm max},i}-m_{{\rm min},i}>20M_{\odot}$} \\
\cline{1-4}
mean of $\chi$ distribution in $i$-th component& $\mu_{{\chi},i}$ &U(0,1) & U(0,1)  \\
standard deviation of $\chi$ distribution in $i$-th component& $\sigma_{{\chi},i}$ &U(0.05, 0.5) & U(0.05, 0.5)  \\
spin constraints & & \multicolumn{2}{c}{$\chi_{{\rm max},i}>\mu_{{\chi},i}>\chi_{{\rm min},i}$} \\
\cline{1-4}
mixing fraction of the $i$-th component& $r_i$ &\multicolumn{2}{c}{{$\mathcal{D}$}}\\
\cline{1-4}
standard deviation of nearly aligned $\cos\theta$ 
& $\sigma_{{\rm t}}$ &\multicolumn{2}{c}{U(0.1, 4)}  \\
mixing fraction of nearly aligned assembly & $\zeta$ &\multicolumn{2}{c}{U(0,1)} \\
\cline{1-4}
pairing function & $\beta$ &\multicolumn{2}{c}{U(0,8)}\\
\cline{1-4}
local merger rate density & $R_0[{\rm Gpc^{-3}yr^{-1}}]$ &\multicolumn{2}{c}{U(0,100)}\\
Hubble constant & $H_0[{\rm km~s^{-1}~Mpc^{-1}}]$  & \multicolumn{2}{c}{U(10,200)} \\
present-day dimensionless matter densities  & $\Omega_{\rm m}$ &  \multicolumn{2}{c}{U(0,1)} \\
slope of the powerlaw regime for the rate evolution before the point $z_{\rm p}$ &$\gamma$&  \multicolumn{2}{c}{U(0,12)} \\
slope of the powerlaw regime for the rate evolution after the point $z_{\rm p}$ &$\kappa$&  \multicolumn{2}{c}{U(0,6)} \\
redshift turning point between the power-law regimes with $\gamma$ and $\kappa$ &$z_{\rm p}$ &  \multicolumn{2}{c}{U(0,4)} 
\end{tabular}

\end{ruledtabular}
\tablenotetext{}{{\bf Note.} Here, U ($\mathcal{N}$) represents the uniform (normal) distribution, and {$\mathcal{D}$ is for the Dirichlet distribution.}
}
\end{table}

\section{Simulation with mock data}\label{sec:sim}

In this section, we conduct simulations using synthetic data to demonstrate the enhancements achieved by multi-spectral sirens compared to traditional spectral sirens. Our analysis primarily targets the lower-edge of the PIMG \citep{2023MNRAS.523.4539K}, a typical feature used in spectral sirens, for measuring cosmological parameters  \citep{2019ApJ...883L..42F, 2022PhRvL.129f1102E}.
Additionally, potential features associated with various subpopulations or formation channels \citep{2022ApJ...941L..39W, 2023arXiv230401288G, 2024arXiv240403166R, 2024arXiv240409668L} may appear in the BH mass function \citep{2023MNRAS.524.5844T, 2024MNRAS.527..298T, 2023ApJ...955..107F}, which could provide more available features and offer further opportunities for multi-spectral sirens.
To clearly illustrate the advantage of multi-spectral sirens, we employ a reduced model to generate the mock population, where the mass functions of the two distinct subpopulations are the simple PowerLaw distributions without the perturbation functions. For the first-generation BHs, we set the parameters as  $m_{{\rm min},1}=5M_{\odot}$, $m_{{\rm max},1}=45M_{\odot}$, $\delta_{{\rm low},1}=5M_{\odot}$, $\delta_{{\rm up},1}=5M_{\odot}$, and $\alpha_1=2.3$. As for the higher-generation BHs, we set the parameters as $m_{{\rm min},2}=30M_{\odot}$, $m_{{\rm max},2}=80M_{\odot}$, $\delta_{{\rm low},2}=0$, $\delta_{{\rm up},2}=0$, and $\alpha_2=2$.
The underlying (unpaired) higher-generation subpopulation takes a fraction of $5\%$, and the pairing function is $\propto(m_2/m_1)^{2.5}$. Note that these configurations are broadly consistent with the results from GWTC-3 \citep{2024PhRvL.133e1401L}, see also the comparison between the (recovered) mass distributions of mock data and the real data in Appendix~\ref{app:mock}.

The spin-magnitude distribution of the first (second) subpopulation is a truncated Gaussian with peak at 0.1 (0.7) and width of 0.15 bounded within (0,1). The spin-orientation distribution for the total BBH population is described by Eq.~(\ref{eq_tilt}). For simplicity, we adopt $\zeta=1$ and $\sigma_{\rm t}=1$, since in this work we do not focus on the subpopulations distinguished by spin-orientation distributions.
The redshift distribution is described by Eq.~(\ref{p_z}), with  parameters $\gamma=2.7$, $\kappa=3$, $z_{\rm p}=2$, $H_0=70 {\rm Mpc}^{-1}~{\rm km}~{\rm s}^{-1}$, and $\Omega_{\rm m}=0.3$. 

We perform simulations subject to the third observing run (O3) of LIGO-Virgo-KAGRA collaboration. In practice, we only incorporate the two advanced LIGO detectors and the advanced Virgo detector since these detectors mainly contribute to the GW detection (see the \href{https://dcc.ligo.org/LIGO-G2002127/public}{Observing Scenario} for details). As for the sensitivity curves, we use the \href{https://dcc.ligo.org/public/0165/T2000012/002/aligo_O3actual_H1.txt}{aligo\_O3actual\_H1}, \href{https://dcc.ligo.org/public/0165/T2000012/002/aligo_O3actual_L1.txt}{aligo\_O3actual\_L1}, and \href{https://dcc.ligo.org/public/0165/T2000012/002/avirgo_O3actual.txt}{avirgo\_O3actual} files, which can be obtained from \href{https://dcc.ligo.org/LIGO-T2000012/public}{LIGO Document Control Center (DCC)} \citep{2018LRR....21....3A}. Besides, we use the Python package \href{https://github.com/CosmoStatGW/gwfast}{GWFast} \citep{2022ApJS..263....2I, 2022ApJ...941..208I} and the {\it IMRPhenomXPHM} waveform \citep{2021PhRvD.103j4056P} to calculate the signal-to-noise ratios (SNRs) for the mock events, and adopt a threshold of network SNR $>11$ to select the mock detections. Here we use a catalog of 50 events for simulation, which is broadly aligned with the GWTC-3. For the parameter estimation of the selected events, we rapidly generate the samples from the corresponding Fisher matrix computed by GWFast \citep{2022ApJS..263....2I, 2022ApJ...941..208I}.

For comparative analysis, we employ two models to recover the mock injections. The first model, referred to as the TwoSpin model, incorporates two subpopulations, each characterized by independent spin-magnitude distributions. The second model, named the NoSpin model, does not differentiate between subpopulations and thus does not model spin distributions. The TwoSpin model has the same formula as Eq.~(\ref{eq:pop}), but employs a simpler component-mass function, i.e., the PowerLaw distributions without the perturbation functions.
To ensure a fair comparison —`apples to apples’— we configure the NoSpin model to use the same mass function formula as the TwoSpin model. Practically, the NoSpin model represents a reduced version of the TwoSpin model, where $\pi_i(\chi)=1$ and $\pi(\cos\theta_1,\cos\theta_2)=1/4$.

\begin{figure}
	\centering  
\includegraphics[width=0.8\linewidth]{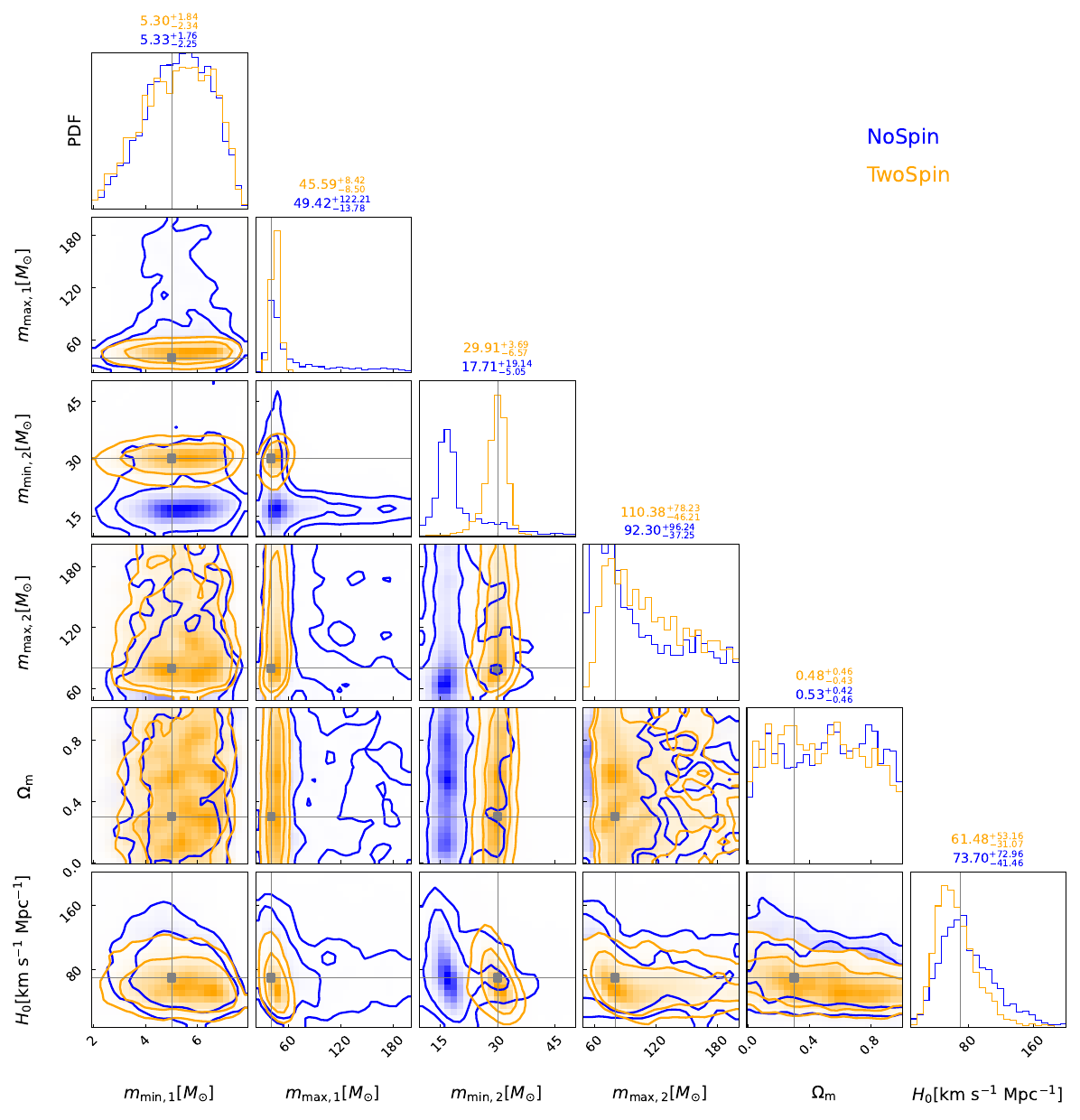}
\caption{Hubble constant, matter density, and the parameters for the population models recovered by the TwoSpin (in orange) and NoSpin (in blue) models with 50 mock detections in O3. The solid lines represent the injections, the contours mark the central 50\% and 90\% posterior credible regions, respectively; the values are for 90\% credible intervals.}
\label{fig:sim_O3_corner}
\end{figure}

\begin{figure}
	\centering
\includegraphics[width=0.6\linewidth]{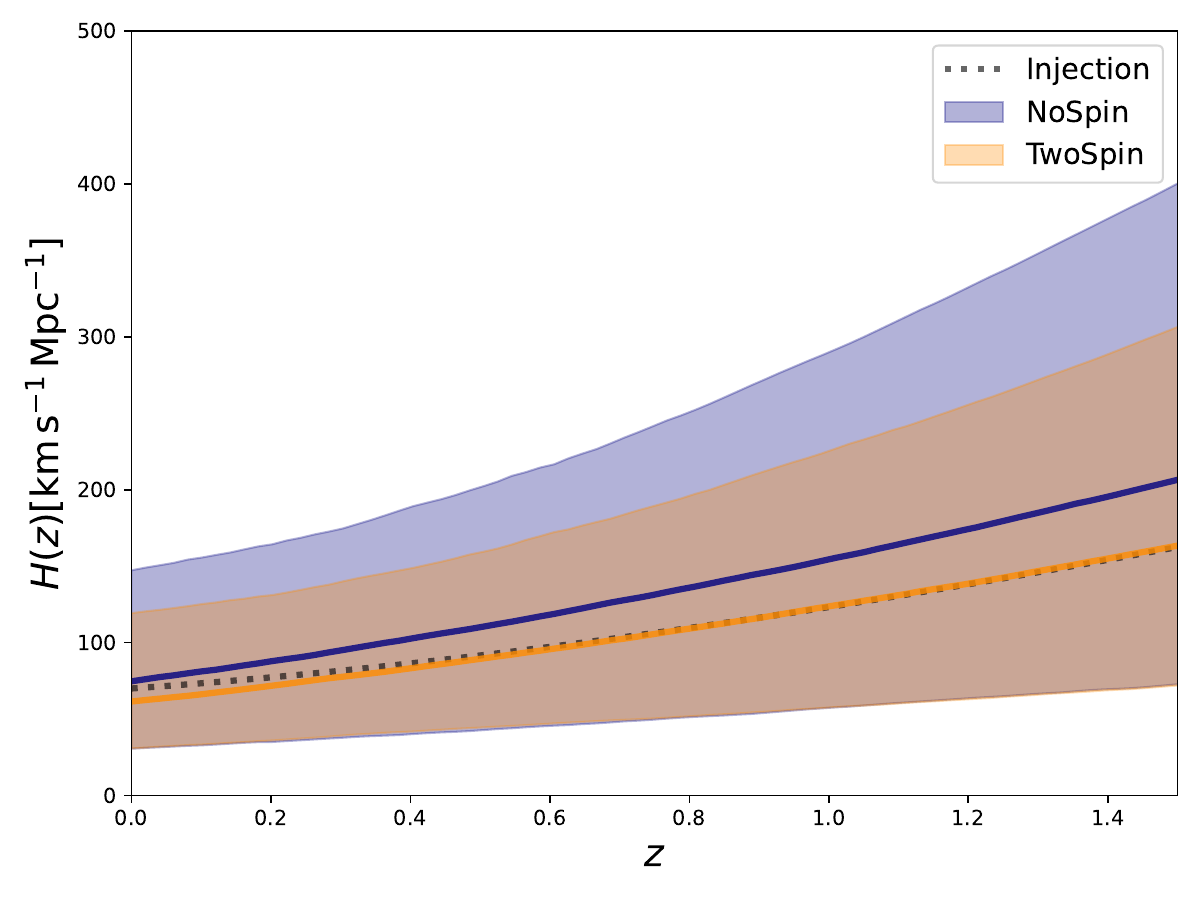}
\caption{Hubble parameters recovered by the TwoSpin (in orange) and NoSpin (in blue) models with 50 mock detections in O3. The doted lines represent the injections. The shaded regions and solid lines are for 90\% credible intervals and median values, respectively.}
\label{fig:sim_O3}
\end{figure}

Figure~\ref{fig:sim_O3_corner} displays the recovered cosmological parameters and population parameters for both models. The results indicate that the Hubble constant ($H_0$) inferred using the TwoSpin model is approximately 20\% more precise than that inferred with the NoSpin model. However, the $\Omega_{\rm m}$ is only weakly constrained by both models.
The improved precision of $H_0$ is attributed to the more effective differentiation of the two subpopulations by the TwoSpin model, which clearly delineates the upper edge of the first subpopulation and the lower edge of the second subpopulation in the multi-spectral sirens, as shown in Figure~\ref{fig:sim_O3_corner}. This distinction also contributes to the accuracy of the $H(z)$ measurement, as shown in Figure~\ref{fig:sim_O3}.
{Within the results of NoSpin model, the minimum mass of the second subpopulation ($M_{\rm min,2}$) is not well recovered and exhibits a bias. This is because, without spin distributions, the NoSpin model is unable to distinguish the two subpopulations, and thus fails to recover $M_{\rm min,2}$ accurately. In contrast, the TwoSpin model has successfully distinguished the two subpopulations and recovered the underlying distributions of the injected mock populations, as shown in Appendix~\ref{app:mock}. Therefore, the inclusion of spin properties can enhance the disentanglement of subpopulations, thereby improving the measurement of cosmological parameters.}

To check the stability of the results, we perform inferences with five independent mock catalogs randomly drawn from the fiducial mock population as described above, and find that all the results are generally consistent across all realizations, see Appendix~\ref{app:check}.

\section{Constraints with GWTC-3}\label{sec:real}
\begin{table*}[htpb]
\centering
\caption{Model comparison}\label{tab:bf}
\begin{tabular}{lcc}
\hline
\hline
Models     &  $\ln{\mathcal{B}}$    \\
\hline
\textsc{TwoSpin}  & 0  \\
PS \& Default Spin & -6.1 \\
PP \& Default Spin & -9.7  \\
\hline
\hline
\end{tabular}
\\
\begin{tabular}{l}
Note: these log Bayes factors are relative to the TwoSpin model in our work.
\end{tabular}
\end{table*} 

Following \citet{2023ApJ...949...76A}, we restrict our main analysis to BBH events with SNRs$>11$, resulting in a total of 42 events. Notably, we exclude GW190814 \citep{2020ApJ...896L..44A} due to its classification as a population outlier \citep{2021ApJ...913L...7A}. We also use a supplementary catalog of BBHs with SNRs$>10$ for further analysis, as illustrated in Appendix~\ref{app:GWTC-3}. It should be noted that the posterior samples used are not cosmologically reweighted, and are available at \href{https://zenodo.org/records/8177023}{zenodo} \citep{https://doi.org/10.5281/zenodo.8177023}. The `Mixed' samples are used for analysis in this work.
The injection campaign data utilized for calculating selection effects also accessible at \href{https://zenodo.org/records/5645777}{zenodo} \citep{the_ligo_2021_5645777}, and the \href{https://zenodo.org/records/5645777/files/O1_O2_O3_det_frame_SNR9.inj}{O1\_O2\_O3\_det\_frame\_SNR9.inj} is adopted. This injection campaign contains injections with detector-frame masses, luminosity distances, and the corresponding drawing probabilities. Therefore, when calculating the selection effects, they should be converted to the source-frame masses, redshifts, and the corresponding drawing probabilities, given the cosmological parameters ($H_0$, $\Omega_{\rm m}$). Note that the drawing probabilities are converted by Jacobian factors of $(1+z)^2/\frac{\partial D_{\rm L}(H_0, \Omega_{\rm m})}{\partial z}$.
We begin by reproducing the results of \citet{2023ApJ...949...76A} with their \textsc{PowerLaw+Peak} (PP) model to ensure the consistency of our framework with theirs. We obtain $H_0=53^{+43}_{-26}~{\rm Mpc}^{-1}~{\rm km}~{\rm s}^{-1}$ at 68.3\% credible level (C.L.), and the posterior distribution is presented in Appendix~\ref{app:PP}, confirming that our reproduced results align with those of \citet{2023ApJ...949...76A}. Note that the population model is not accompanied with spin distributions.
 
We use the Eq.~(\ref{eq:pop}) as the multi-spectral sirens model to jointly infer the cosmological parameters and BBH population. Following \citet{2024PhRvL.133e1401L}, we choose a two-component mixture model for Eq.~(\ref{eq:mix}) \footnote{It is sufficient for currently available data, and a more complex mixture model may be needed when the data are significantly enriched, see \citet{2024PhRvL.133e1401L} for detailed illustration.}.
This is a flexible and semi-parametric model, which can determine the underlying mass functions of the subpopulations with minimal prior assumptions for the shape of the mass distributions, so that can mitigate the inaccurate representation that leads to biases in the cosmological inference \citep{2024PhRvD.109h3504P}.
We obtain $H_0=73.3^{+29.9}_{-25.6} (^{+51.7}_{-40.6})~{\rm Mpc}^{-1}~{\rm km}~{\rm s}^{-1}$ at 68.3\% (90\%) C.L. for the TwoSpin model, which is $\sim 19\%$ tighter than that inferred with PP model.

\begin{figure}
	\centering  
\includegraphics[width=0.48\linewidth]{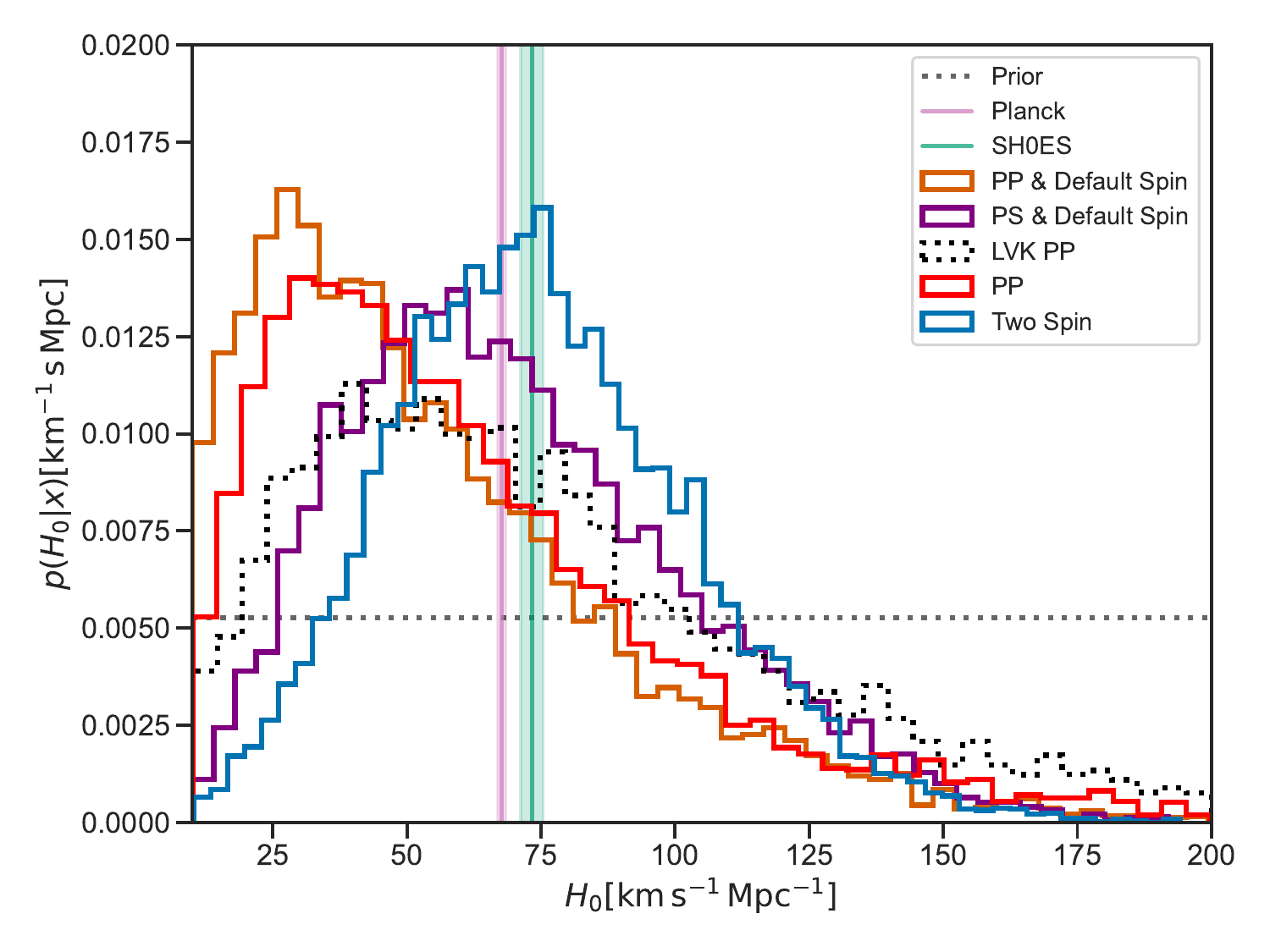}
\includegraphics[width=0.48\linewidth]{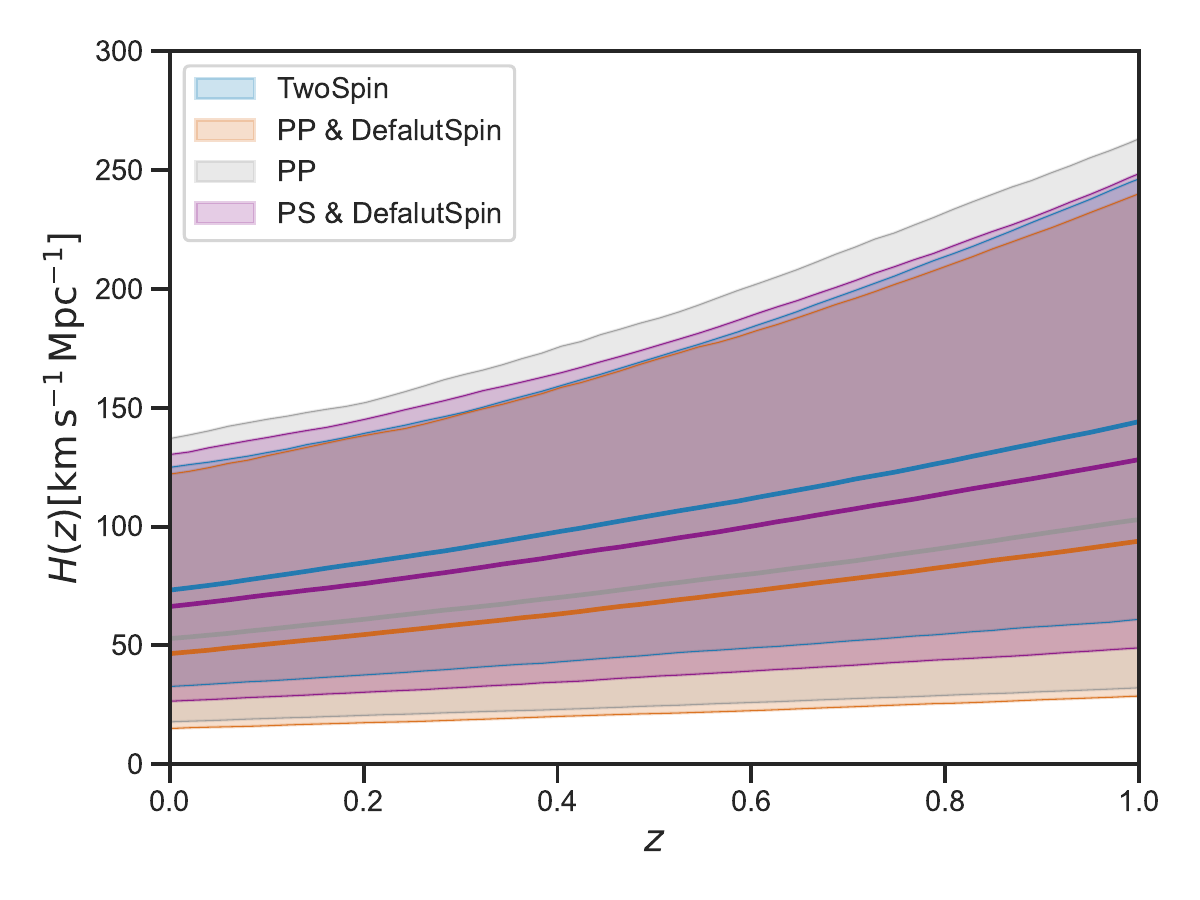}
\caption{The Hubble constant (left) and expansion rate (right) inferred using multi-spectral sirens (i.e., the TwoSpin model) and the traditional spectral sirens (PP and PS models), with 42 events (SNRs$>11$). In the left panel, the pink and green bands represent the Hubble constant measured from CMB \citep{2020A&A...641A...6P} and that measured in the local universe \citep{2019ApJ...876...85R}, respectively. The doted histogram is the results adopted from \cite{2023ApJ...949...76A}, and the doted line represent the prior distribution.
In the right panel, the shaded regions and solid lines are for 90\% credible intervals and median values, respectively.}
\label{fig:H0_comp}
\end{figure}

For comparison, we also adopt the \textsc{PowerLaw+Spline} (PS) model \citep{2022ApJ...924..101E,2023PhRvX..13a1048A} and PP model, accompanied with the \textsc{Default} Spin model \citep{2023PhRvX..13a1048A}. In Table~\ref{tab:bf}, we report the Bayes factors computed between different population models. We find that the TwoSpin model is significantly more favored than the PP and PS models (with Default Spin model) by Bayes factors of $\ln{\mathcal{B}}=9.7$ and $\ln{\mathcal{B}}=6.1$, respectively. This preference arises from the TwoSpin model's superior ability to characterize the two subpopulations within the component-mass versus spin-magnitude distribution compared to the PP and PS models \citep[see][for a detailed illustration]{2024PhRvL.133e1401L}. Such a result is consistent with the fact that the multi-spectral sirens has provided tighter constraints of the cosmological parameters than the traditional spectral sirens, as show in figure~\ref{fig:H0_comp}. The PS model is more preferred than the PP by a Bayes factor of $\ln{\mathcal{B}}=3.6$. This is because the PS model is more flexible than the PP model thus better modeling the mass function of BBHs \citep[see][for detailed illustration]{2022ApJ...924..101E}.

Figure~\ref{fig:H0_comp} (left) shows the Hubble constants inferred with multi-spectral sirens (i.e., TwoSpin model) and traditional spectral sirens (i.e., PP and PS models),  
which are broadly consistent with Planck \citep{2020A&A...641A...6P} and SH0ES \citep{2019ApJ...876...85R} measurements . We have also included the inferred $H_0$ of PP model in \citet{2023ApJ...949...76A} \footnote{The posterior samples are adopted from \url{https://zenodo.org/records/5645777}.} for comparison, which are slightly less constrained than our results. Note that this may partially attribute to the difference of the assumed cosmological models between this work and \citet{2023ApJ...949...76A}. We use a Flat$\Lambda$CDM model, whereas \citet{2023ApJ...949...76A} use a Flat$w_0$CDM model \footnote{{Note that \citet{2023ApJ...949...76A} have also used a Flat$\Lambda$CDM model with fixed values of $\Omega_{\rm m}= 0.3065$ and $w_0=-1$. However, the prior for $H_0$ is restricted to [65, 77] ${\rm Mpc}^{-1}~{\rm km}~{\rm s}^{-1}$ in that model, which is not suitable for comparison in this work.}}.
For the expansion rate $H(z)$, we find the TwoSpin model provides tighter constraint than the other models, which is consistent with the case using events with SNR$>10$. Additionally, we find the PP model accompanied with the \textsc{Default} spin model provides a slightly tighter constraint than the PP model without spin distribution; however, systematic errors might have been included by the mis-modeling of spin distribution by the \textsc{Default} spin model, as indicated by the Bayes factors shown in Table~\ref{tab:bf}. 

With the multi-spectral sirens model, we simultaneously identified two subpopulations of BHs, similar to those found in \citet{2024PhRvL.133e1401L}, see Appendix~\ref{app:GWTC-3}.
The maximum mass of the first subpopulation (i.e., constrained to $\sim40M_{\odot}$) has contributed to the measurement of cosmology parameters, which is potentially associated with the lower edge of PIMG \citep{2024PhRvL.133e1401L}. This feature is not incorporated into the single-population model \citep[e.g.][]{2023PhRvX..13a1048A}, hence can not contribute to traditional spectral sirens \citep{2023ApJ...949...76A,2024arXiv240402210F,2024arXiv240402522M}.
We note the $m_{{\rm max},1}$ has a tail extend to high-mass range, which is resulted from the flexibility of the \textsc{PowerLawSpline} mass function (see also \citet{2024PhRvL.133e1401L} for illustration). However the mass of 99\% and 99.5\% for the first subpopulation is better measured and more degenerated with $H_0$, Appendix~\ref{app:GWTC-3}.

In addition to the maximum mass of the first subpopulation $m_{{\rm max},1}$ (or $m_{99\% , 1}$,  $m_{99.5\% , 1}$), we observe that the $m_{{\rm min},2}$ and $m_{{\rm max},2}$ are also degenerate with the Hubble constant, see Appendix~\ref{app:GWTC-3}. 
Notably, the minimum mass of the second subpopulation ($m_{{\rm min},2}$) represents an additional feature of multi-spectral sirens compared to traditional spectral sirens, and is anticipated to enhance the measurement of cosmological parameters, as discussed in Section~\ref{sec:sim}. Furthermore, the rate evolution parameter $\gamma$ exhibits a correlation with the estimation of $H_0$, a relationship that remains consistent across both the TwoSpin model and the PP model; specifically, higher values of $\gamma$ are associated with lower values of $H_0$ \citep[see also][]{2023ApJ...949...76A}.

\begin{figure}
	\centering  
\includegraphics[width=0.6\linewidth]{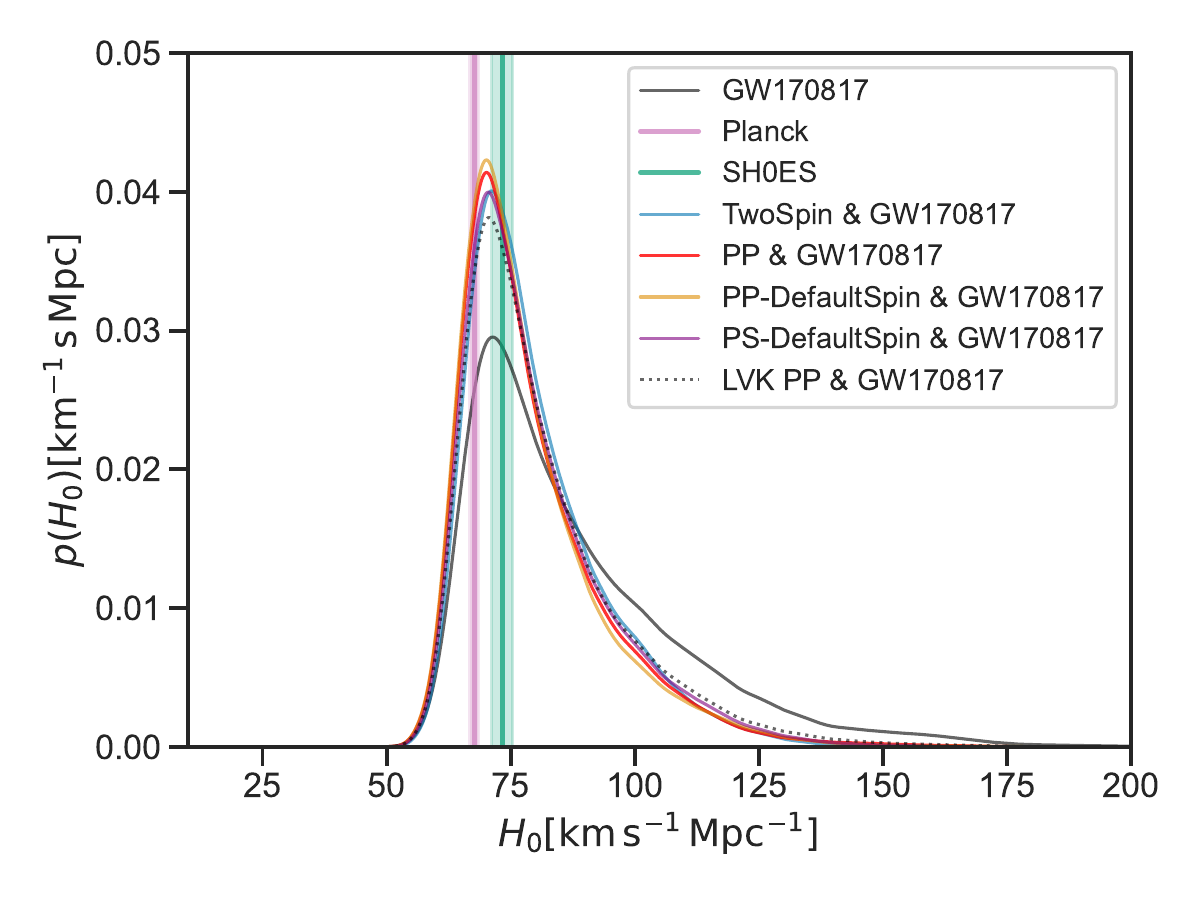}
\caption{Posterior distributions for $H_0$ obtained by combining the measurements from multi-spectral sirens / spectral sirens with 42 events (SNR$>11$) and measurement from the bright standard siren GW170817 \citep{2017Natur.551...85A}.}
\label{fig:H0_GW170817}
\end{figure}

We also combine the $H_0$ posteriors from the multi-spectral sirens and spectral sirens with the $H_0$ inferred from the bright standard siren GW170817 \footnote{The posterior samples of the $H_0$ is directly adopted from \url{https://dcc.ligo.org/public/0145/P1700296/005/ExtendedDataFigure2.csv}.} \citep{2017Natur.551...85A} (see Figure~\ref{fig:H0_GW170817}), and obtain $H_0=71.1^{+15.0}_{-7.5}~{\rm Mpc}^{-1}~{\rm km}~{\rm s}^{-1}$ ($H_0=70.3^{+12.9}_{-7.1}~{\rm Mpc}^{-1}~{\rm km}~{\rm s}^{-1}$) with a uniform prior in $[10,200]~{\rm Mpc}^{-1}~{\rm km}~{\rm s}^{-1}$ (log-uniform prior in $[20,140]~{\rm Mpc}^{-1}~{\rm km}~{\rm s}^{-1}$) at 68.3\% HDI for the TwoSpin model. This combined measurement is about $26\%$ ($23\%$) better than the measurement from bright standard siren GW170817 alone \citep{2017Natur.551...85A}.
The results of the PP models, both with and without spin distribution, combined with GW170817, yield slightly tighter constraints compared to those derived from the TwoSpin model combined with GW170817. However, the opposite trend is observed when considering events with SNR$>$10.
Specifically, for events with SNR$>$11, the $H_0$ measurements from the PP models exhibit less support for larger values (e.g., $H_0>100~{\rm Mpc}^{-1}~{\rm km}~{\rm s}^{-1}$) compared to the TwoSpin model. In contrast, GW170817 provides a constraint on $H_0$ that excludes smaller values, resulting in a tighter constraint when combining the PP models with GW170817. Conversely, for events with SNR$>$10, the TwoSpin model demonstrates a distribution of $H_0$ that shows less support for larger values than the other models, thereby facilitating a tighter constraint on $H_0$ when combined with GW170817.

\section{Summary and Discussion}\label{sec:dis}
We constrain the parameters of the cosmological model with mass functions of subpopulations of BBHs/BHs (i.e., via multi-spectral sirens), and illustrate the advantage of the multi-spectral sirens with a simulation study. The mock population consists of two subpopulations with different spin-magnitude distributions, as motivated by the findings with GWTC-3 data \citep{2024PhRvL.133e1401L}. 
With the identification of the second subpopulation (higher-generation BHs), the first subpopulation (first-generation BHs) presents a more clear cut-off/drop, 
which can provide a better measurement of cosmological parameters \citep{2019ApJ...883L..42F, 2022PhRvL.129f1102E}, since $m_{\rm max,1}$ and $m_{\rm min,2}$ are correlated with $H_0$, as shown in Figure~\ref{fig:sim_O3_corner}.
Then we apply this novel multi-spectral sirens to the real data analysis with GWTC-3 \citep{2023ApJ...949...76A}, and obtain $H_0=73.3^{+29.9}_{-25.6} (^{+51.7}_{-40.6})~{\rm Mpc}^{-1}~{\rm km}~{\rm s}^{-1}$ (with the TwoSpin model) at 68.3\% (90\%) C.L. 
Further analysis, incorporating the bright standard siren GW170817 \citep{2017Natur.551...85A} with a uniform prior in [10, 200] (log-uniform prior in [20,140]) ${\rm Mpc}^{-1}~{\rm km}~{\rm s}^{-1}$, improved the precision of our estimate to $H_0=71.1^{+15.0}_{-7.5}~(70.3^{+12.9}_{-7.1})~{\rm Mpc}^{-1}~{\rm km}~{\rm s}^{-1}$ at the 68.3\% HDI, which represents a 26\% (23\%) enhancement in precision over the measurement derived solely from GW170817.
 
In our mock data study, we only focus on the constraint on the Hubble constant attribute to the maximum (minimum) mass of the first (second) subpopulation.
In reality, the features could be more complex. For example, the mass distribution will evolve with redshift since the metallicity of BBH progenitor systems changes with the age of the universe \citep{2017MNRAS.472.2422M,2021MNRAS.504..146V}. Such evolution, however, can be simultaneously constrained, unless all features in the full mass distribution evolve with the same behavior \citep{2022PhRvL.129f1102E}.
Moreover, additional features in the (subpopulation) mass functions may present, like the peaks/bumps and gaps \citep{2021ApJ...917...33L, 2023MNRAS.524.5844T, 2023ApJ...955..107F, 2022ApJ...928..155T, 2024MNRAS.527..298T, 2024PhRvX..14b1005C}.
The previously found excesses at around $10M_{\odot}$ and $32M_{\odot}$ \citep{2023PhRvX..13a1048A} may be associated with diverse subpopulations originated from the isolated field evolution and dynamical formation channels, respectively \citep{2023arXiv230401288G, 2024arXiv240403166R, 2024arXiv240409668L}. The identification of these subpopulations will also potentially enhance the advantage of the multi-spectral sirens. However, there is still vast uncertainty with currently available data \citep[see e.g.,][]{2023arXiv230401288G,2024arXiv240409668L}.
\citet{2022PhRvL.129f1102E} suggested that the expected low-mass gap \citep[LMG,][]{1974PhRvL..32..324R,1996ApJ...470L..61K,1998ApJ...499..367B,2010ApJ...725.1918O,2011ApJ...741..103F} will play a dominant role on spectral sirens in era of next-generation GW detectors. However, the LMG is now found populated with unknown-origin sources \citep{2020ApJ...896L..44A, 2024arXiv240404248T}. It is possible that the events in the LMG, may suffer from some unconventional evolution paths, e.g., the hierarchical mergers \citep{2020PhRvD.101j3036G} and/or super-Eddington accretion \citep{2022MNRAS.514.1054G}, which is identifiable by their distinctive spin properties.
In our analysis, the spin-magnitude distributions are adequately represented using simple Gaussian models, which suffice given the current data \citep[see Supplemental Material of][for detailed ilustration]{2024PhRvL.133e1401L}. However, more sophisticated models, such as non-parametric ones including auto-regressive processes \citep{2024PhRvX..14b1005C} and splines \citep{2023PhRvD.108j3009G, 2023arXiv230401288G}, could be necessary as data sets become more comprehensive. Additionally, we have not differentiated between redshift distributions across various subpopulations due to the limited data presently available. This aspect will be included in the analysis of future data \citep{2018LRR....21....3A}.

Note that including the spin parameters may also influence the mass distribution \citep[see e.g.,][]{2021ApJ...921L..15G}, so that may influence the measurement of the cosmological parameters in the Spectral Sirens. However, we find that such an effect is weaker than the enhancement introduced by the multi-Spectral Sirens, where the spin parameters are mainly used for splitting the mass distribution into subpopulations (see Figure~\ref{fig:H0_comp}). Additionally, mis-modeling the spin versus mass distribution might bring systemic errors. Therefore, in this work, we can not conclude that including spin parameters in the traditional Spectral Siren has made significant improvement in the measurement of the cosmological parameters.

In the final stage of this work, there was a work appeared online \citep{2024arXiv240606109U}, where the authors investigate the generations in the BBH populations, and then apply their population model to constrain the Hubble constant with GWTC-3. Different from the results in this work, as well as our previous works \citep{2022ApJ...941L..39W, 2024PhRvL.133e1401L, 2024arXiv240409668L}, they report the lower-edge of the PIMG to be $\sim 84 M_{\odot}$ and $H_0 \sim 36 ~{\rm Mpc}^{-1}~{\rm km}~{\rm s}^{-1}$. These discrepancies may be driven by the difference in the mass functions between our models and theirs.
  
{\bf Data availability:} The codes used for this work are publicly available in \href{https://github.com/JackLee0214/Multi_spectral_sirens}{GitHub}.

\begin{acknowledgments}
We thank Yi-Ying Wang and Bo Gao for helpful discussion.
This work is supported by the National Natural Science Foundation of China (No. 12233011, No. 12203101 and No. 12303056), the Priority Research Program of the Chinese Academy of Sciences (No. XDB0550400), and the General Fund (No. 2023M733736, No. 2024M753495) and the Postdoctoral Fellowship Program (GZB20230839) of the China Postdoctoral Science Foundation. This research has made use of data and software obtained from the Gravitational Wave Open Science Center (https://www.gw-openscience.org), a service of LIGO Laboratory, the LIGO Scientific Collaboration and the Virgo Collaboration. LIGO is funded by the U.S. National Science Foundation. Virgo is funded by the French Centre National de Recherche Scientifique (CNRS), the Italian Istituto Nazionale della Fisica Nucleare (INFN) and the Dutch Nikhef, with contributions by Polish and Hungarian institutes.
\end{acknowledgments}

\vspace{5mm}

\software{Bilby \citep[version 1.1.4, ascl:1901.011, \url{https://git.ligo.org/lscsoft/bilby/}]{2019ascl.soft01011A},
          PyMultiNest \citep[version 2.11, ascl:1606.005, \url{https://github.com/JohannesBuchner/PyMultiNest}]{2016ascl.soft06005B},
          Icarogw \citep[\url{https://git.ligo.org/cbc-cosmo/icarogw}]{2021PhRvD.104f2009M,2023arXiv230517973M}.
          }

\appendix

\section{Check the consistency}\label{app:check}

\begin{figure}
	\centering  
\includegraphics[width=0.8\linewidth]{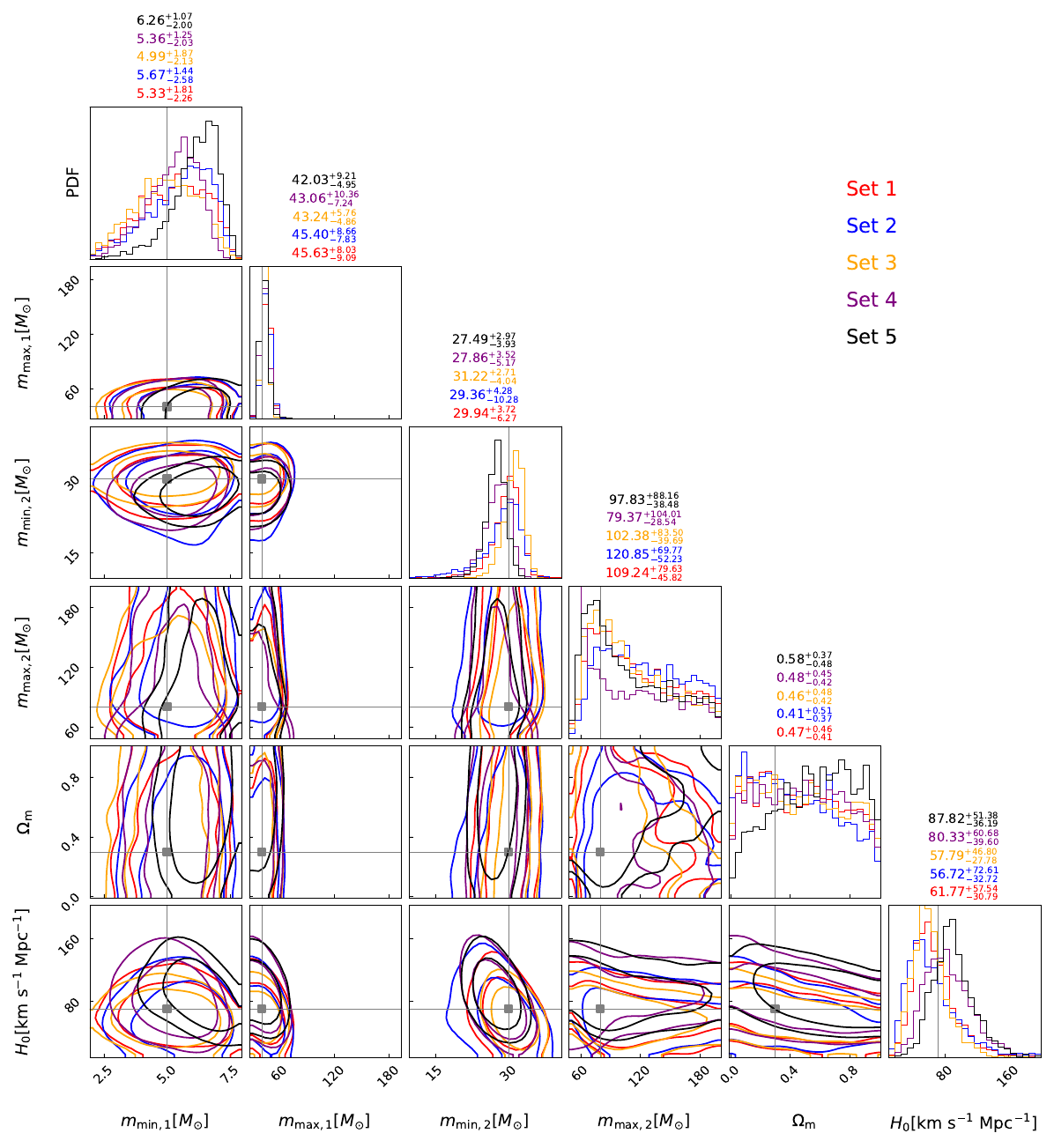}
\caption{Hubble constant, matter density, and the parameters for the population models recovered by the TwoSpin model with 5 independent sets of mock detections in O3. The solid lines represent the injections, the contours mark the central 50\% and 90\% posterior credible regions, respectively; the values are for 90\% credible intervals.}
\label{fig:consistency}
\end{figure}

To check the stability of the inferred results in the mock data simulations, we perform inference with 5 independent mock catalogs. Each catalog includes 50 detections generated by the same procedure as introduced in Section~\ref{sec:sim}. Figure~\ref{fig:consistency} displays the main parameters of population model and the cosmological parameters recovered by the TwoSpin model with the 5 independent catalogs. It shows that all the results are generally consistent with each other.

\section{Additional results of TwoSpin model}

\subsection{With data of GWTC-3}\label{app:GWTC-3}
\begin{figure}
	\centering  
\includegraphics[width=0.96\linewidth]{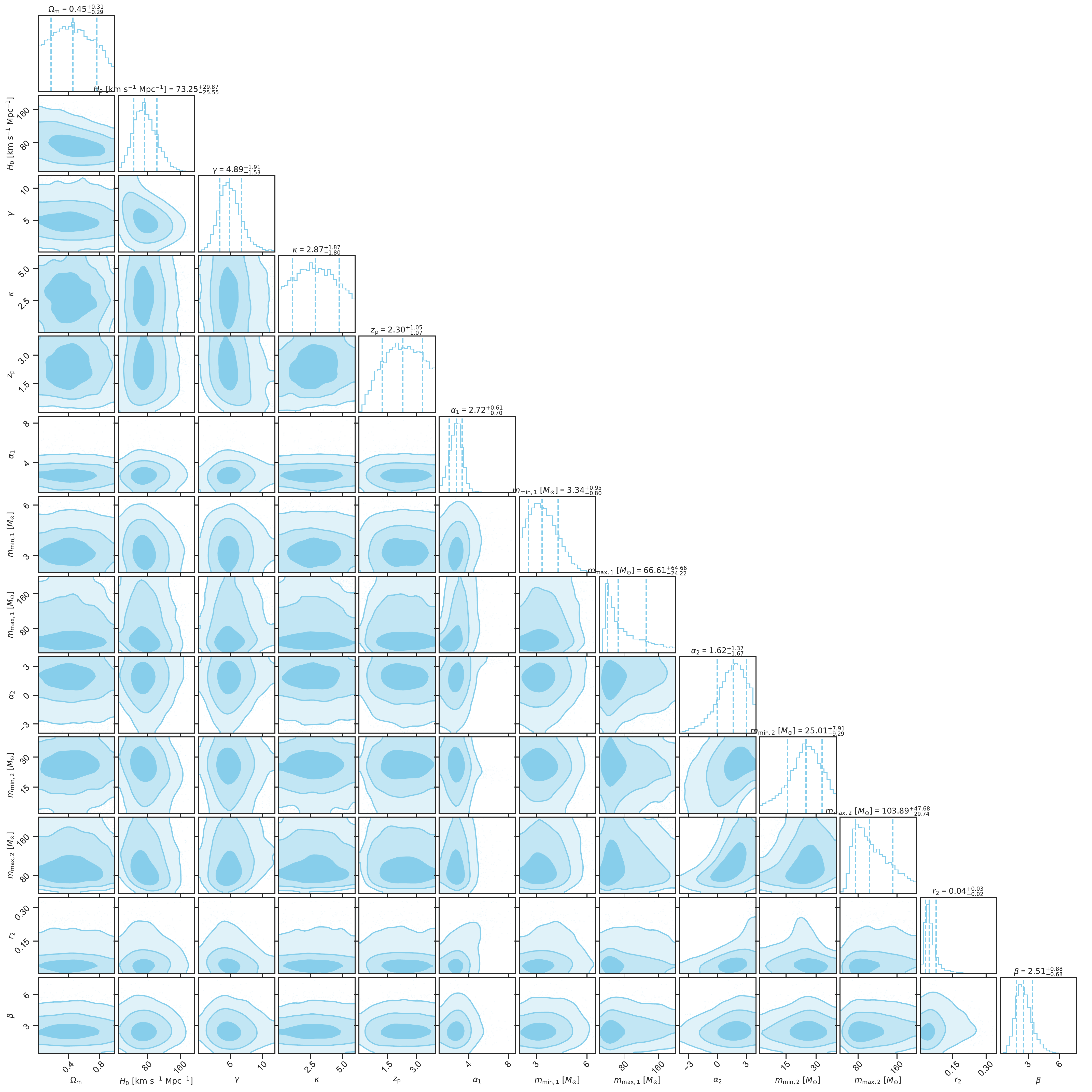} \caption{Posterior distributions of the hyper-parameters inferred with the TwoSpin model and events in GWTC-3 selected by SNR$>11$. The dashed lines in the marginal distribution represent the median and 68.3\% credible intervals.}
\label{fig:TwoSpincorner}
\end{figure}

\begin{figure}
	\centering  
\includegraphics[width=0.6\linewidth]{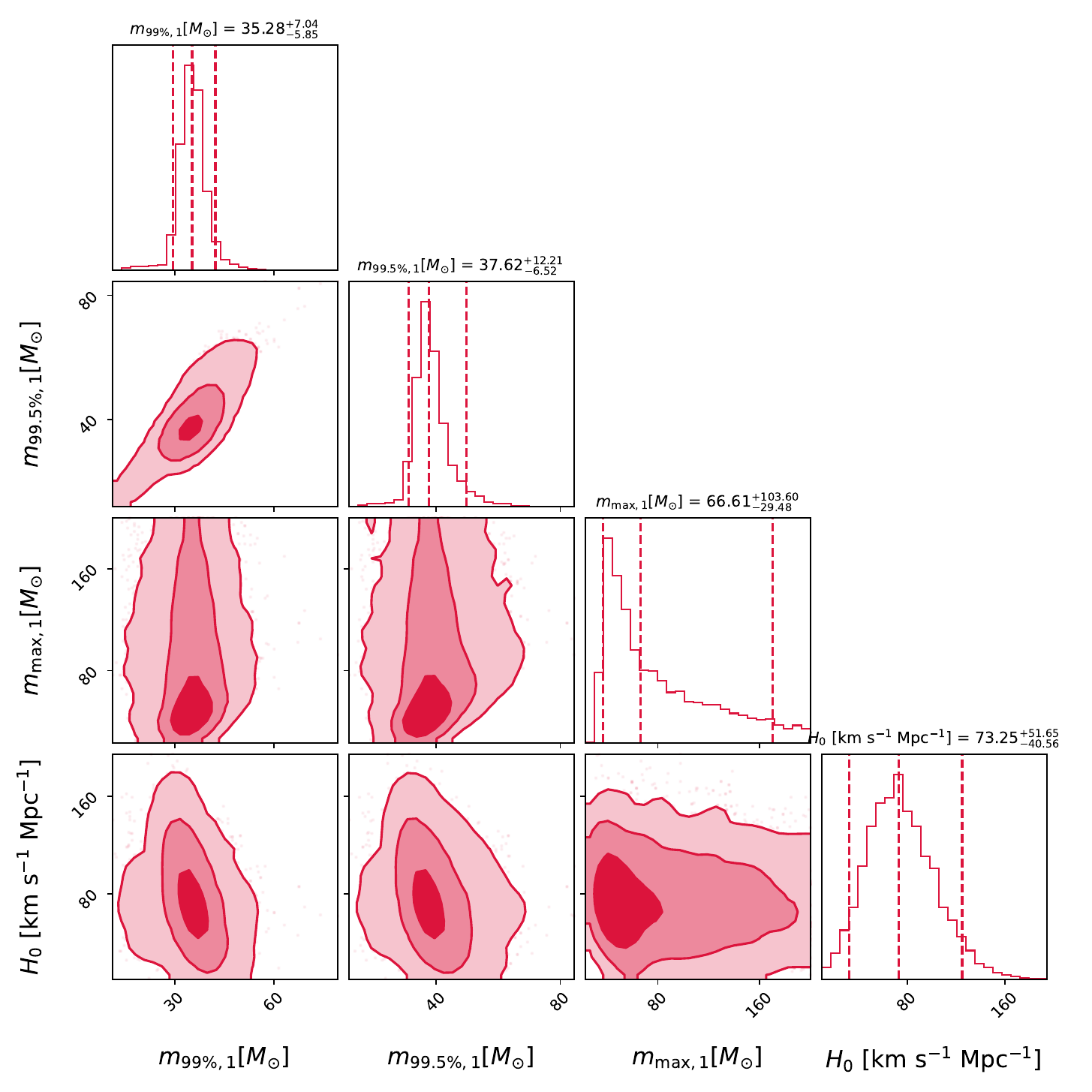}
\caption{Distributions of the Hubble constant, the maximum mass of the first subpopulation, and the masses of the 99.5\% and 99\% percentiles. The dashed lines in the marginal distribution represent the median and 90\% credible intervals.}
\label{fig:corner}
\end{figure}

Figure~\ref{fig:TwoSpincorner} displays the posterior distributions of the hyper-parameters inferred by the TwoSpin model using 42 BBHs with SNRs$>11$ in GWTC-3. It shows that both $m_{{\rm max},1}$ and $m_{{\rm max},2}$ are correlated with the Hubble constant. Though the maximum mass of the first-generation subpopulation has a tail extend to the higher-mass range, the masses of the 99th and 99.5th percentiles are better constrained and clearly correlated with the Hubble constant, as shown in Figure~\ref{fig:corner}.

Figure~\ref{fig:TwoSpin_dist} shows the inferred mass versus spin-magnitude distributions of the two subpopulations using 42 BBHs with SNRs$>11$ in GWTC-3, which are consistent with the results obtained with a fixed cosmological model \citep{2024PhRvL.133e1401L}.

\begin{figure}
	\centering  
\includegraphics[width=0.32\linewidth]{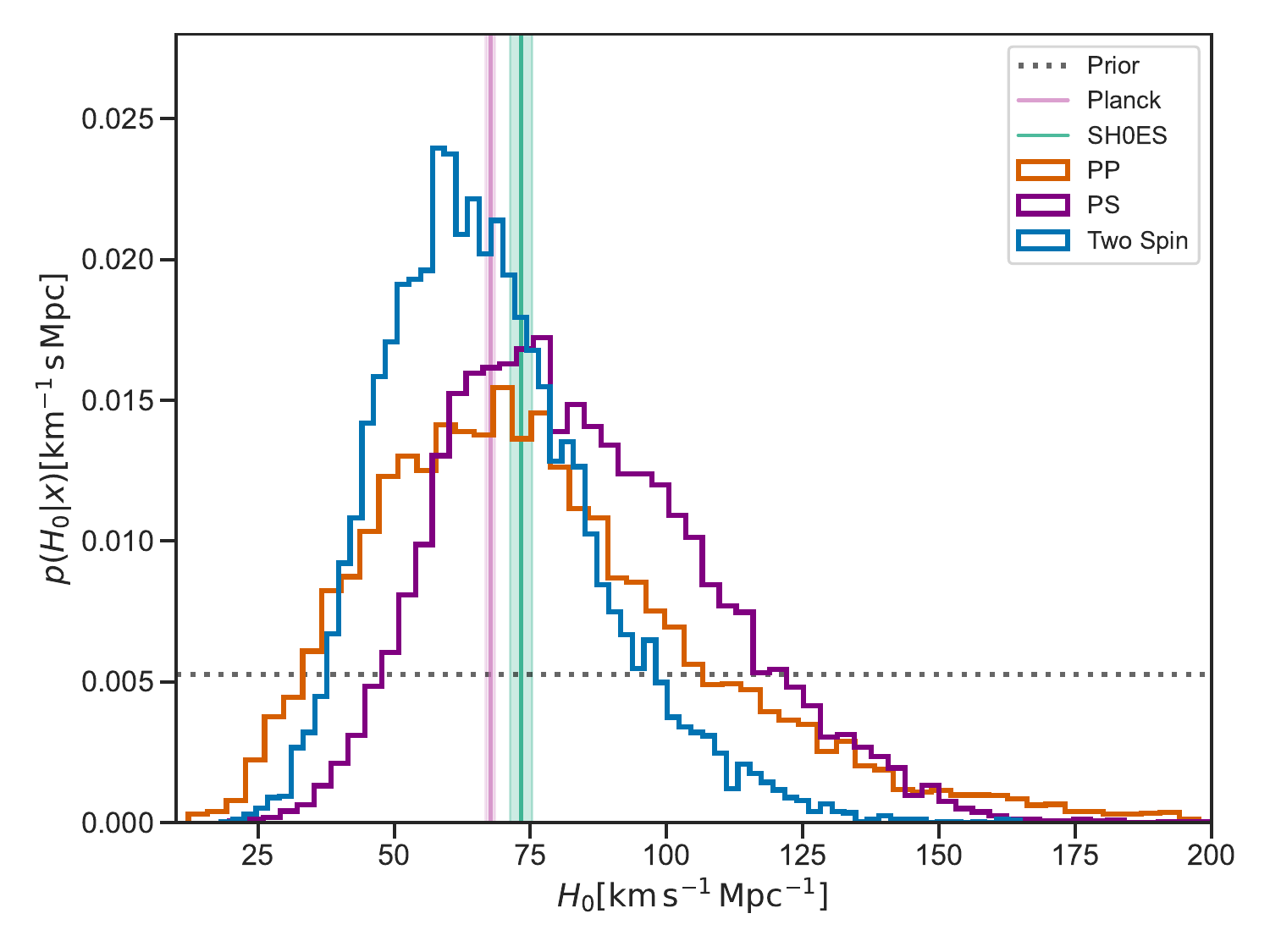}
\includegraphics[width=0.32\linewidth]{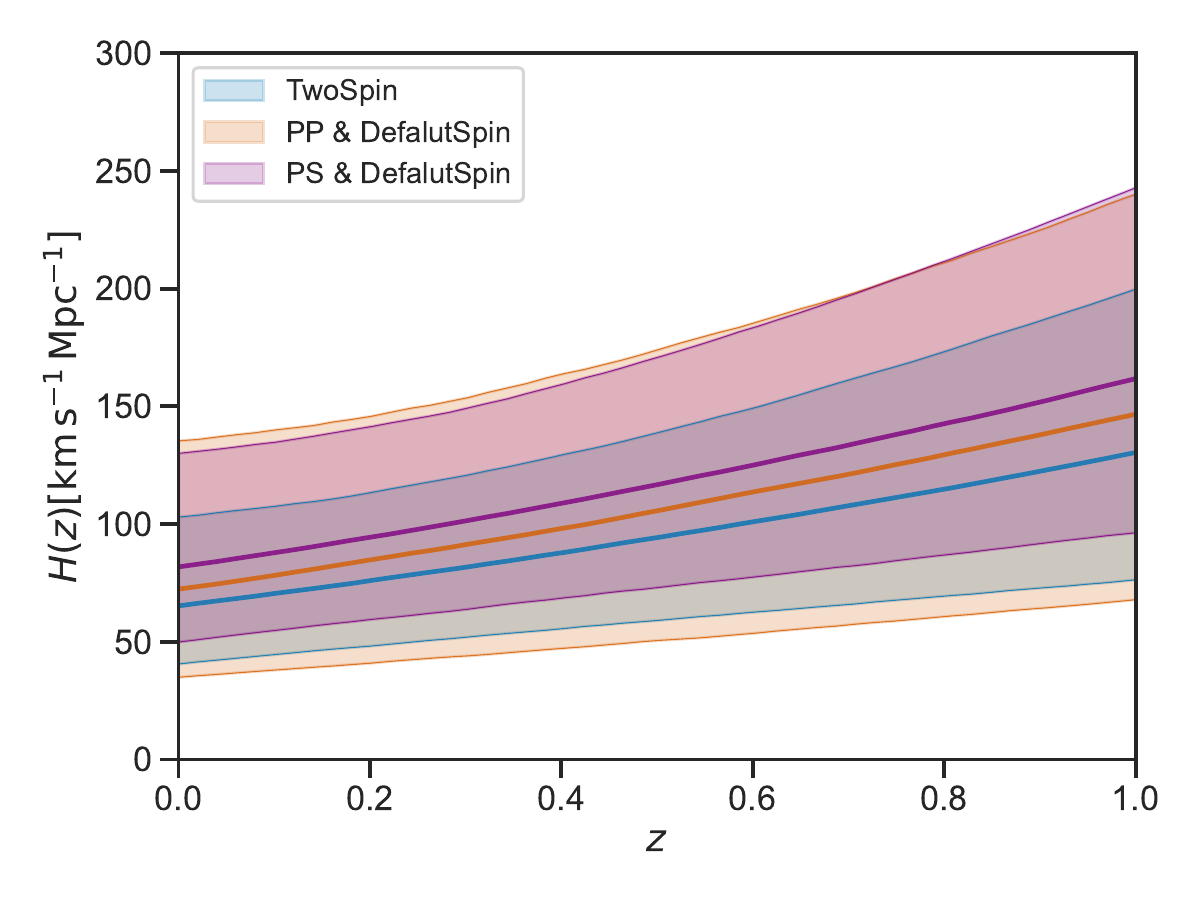}
\includegraphics[width=0.32\linewidth]{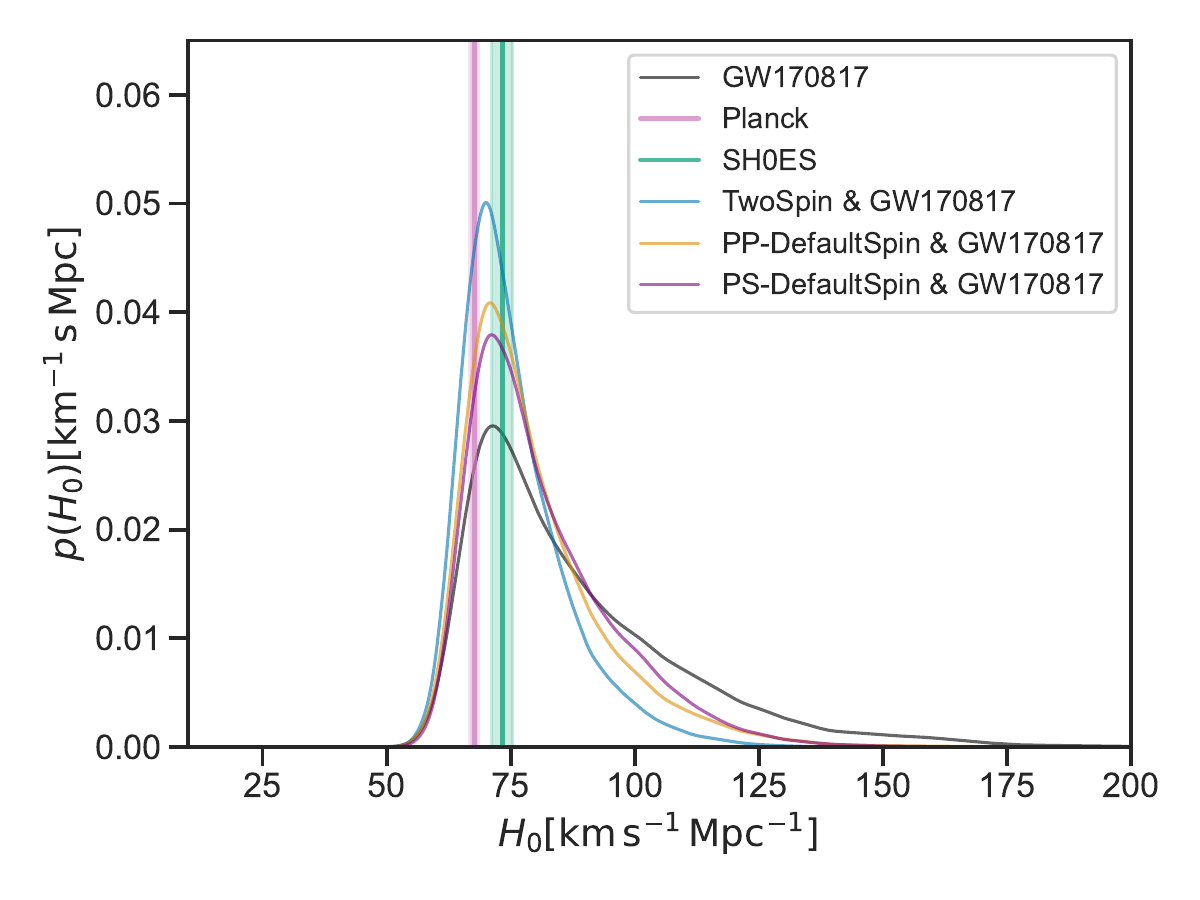}
\caption{Hubble constant (Left) and expansion rate (Mid) inferred with events in GWTC-3 selected by SNR$> 10$, and the results combined with GW170817 \citep{2017Natur.551...85A} (Right). In the left and right panels, the pink and green bands represent the Hubble constant measured from CMB \citep{2016A&A...594A..13P} and that measured in the local universe \citep{2019ApJ...876...85R}, respectively. The doted line represent the prior distribution.
In the mid panel, the shaded regions and solid lines are for 90\% credible intervals and median values, respectively.}
\label{fig:H0_comp_snr_10}
\end{figure}

Following \cite{2023ApJ...949...76A}, we have also performed inference with events selected by SNR$>10$, see Figure~\ref{fig:H0_comp_snr_10} for the results. It shows that the multi-spectral siren (i.e., TwoSpin model) gives more precise Hubble constant and expansion rate than the traditional spectral sirens (i.e., PP and PS model).

\subsection{With mock data}\label{app:mock}
\begin{figure}
	\centering  
\includegraphics[width=0.96\linewidth]{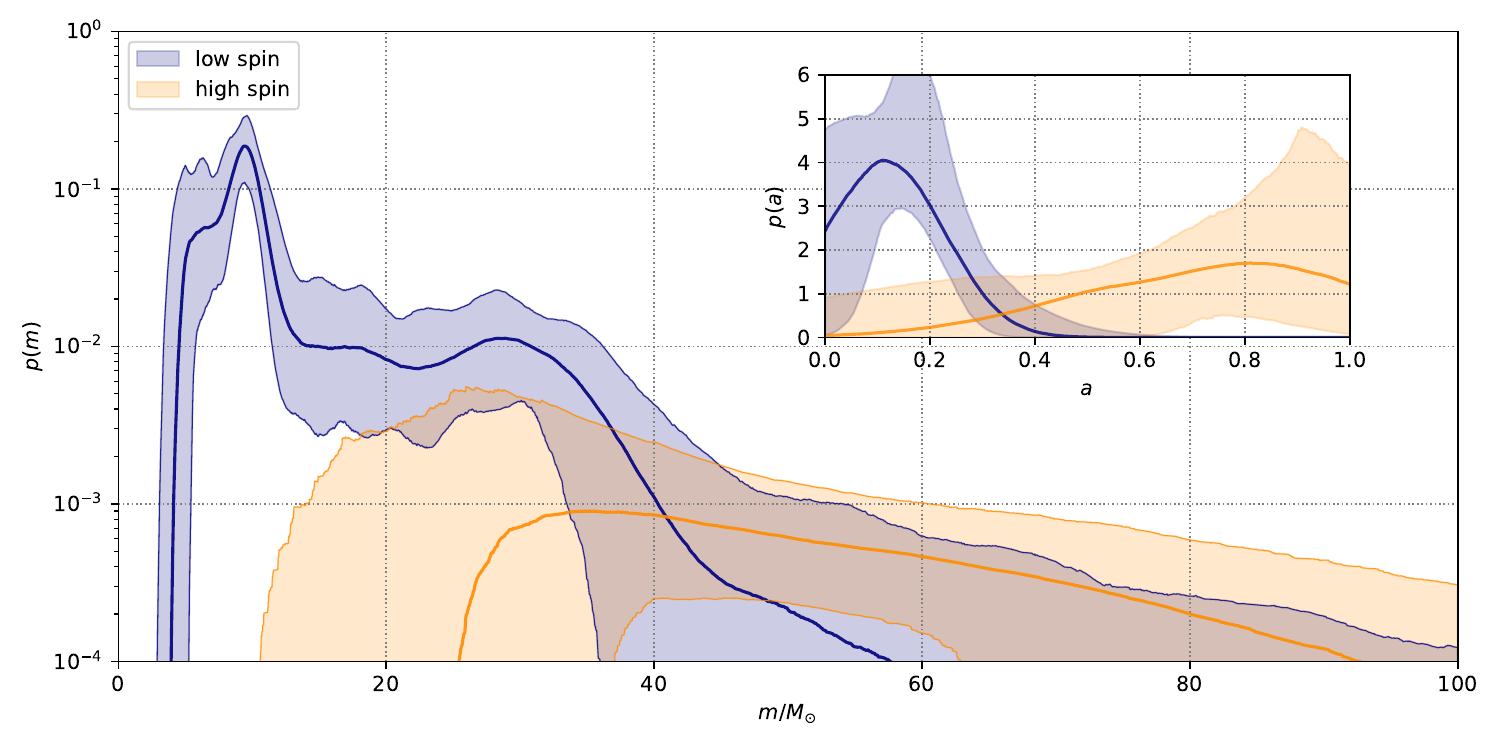}
\caption{Component-mass and spin-magnitude distributions in the multi-spectral sirens, inferred with the TwoSpin model and 42 BBHs with SNRs $>11$ in GWTC-3. The solid lines and dashed regions are for the median and 90\% credible levels.}
\label{fig:TwoSpin_dist}
\end{figure}

\begin{figure}
	\centering  
\includegraphics[width=0.96\linewidth]{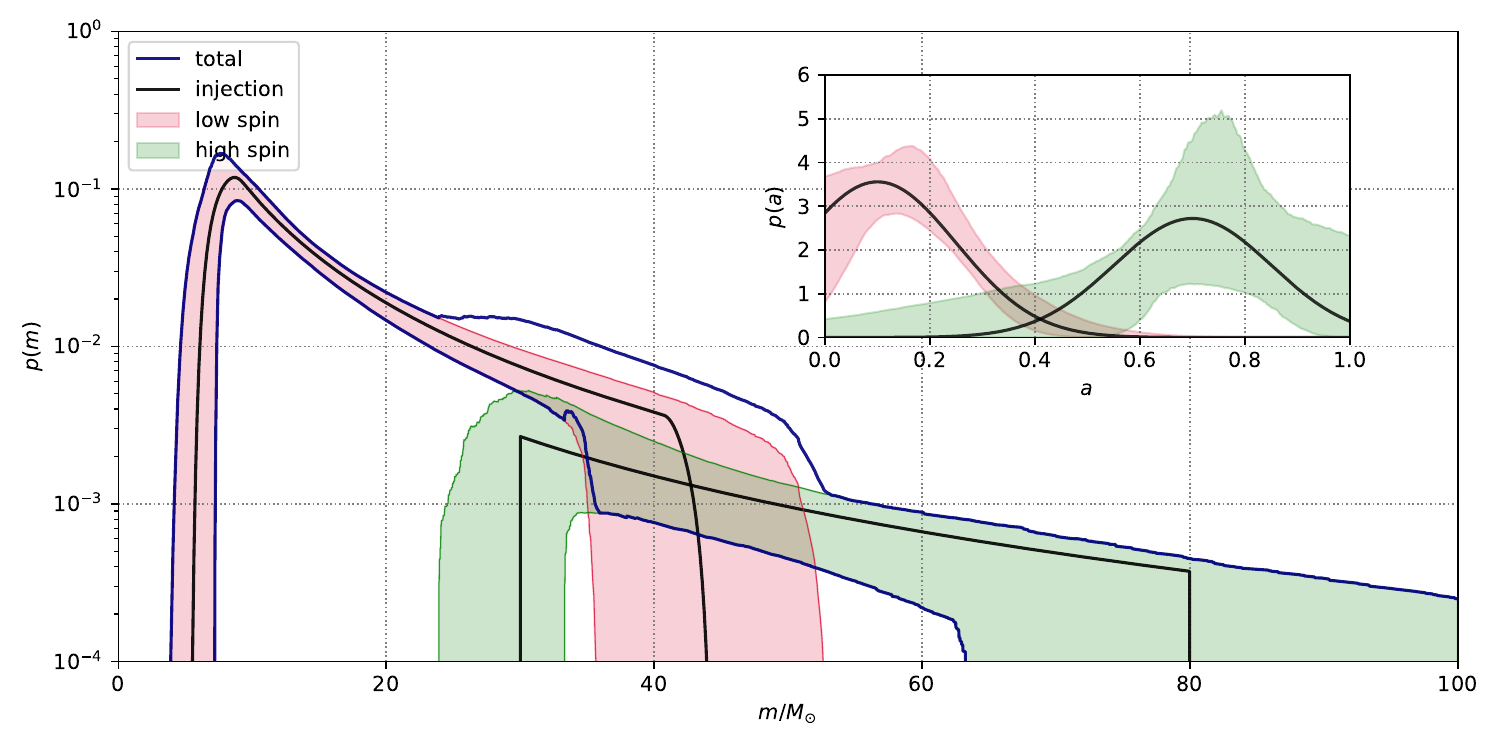}
\caption{Component-mass and spin-magnitude distributions of the subpopulations in the multi-spectral sirens, inferred using the TwoSpin model with 50 mock detections in O3. The solid lines and dashed regions are for the median and 90\% credible levels.}
\label{fig:sim_dist}
\end{figure}

Figure~\ref{fig:sim_dist} shows the mass versus spin-magnitude distributions of the two subpopulations recovered with the mock population in the Section~\ref{sec:sim}, which are broadly consistent with the results inferred with the real data, as shown in Figure~\ref{fig:TwoSpin_dist}, see also \cite{2024PhRvL.133e1401L}.

\section{Result with \textsc{PowerLaw+Peak} model}\label{app:PP}
\begin{figure}
	\centering  
\includegraphics[width=0.96\linewidth]{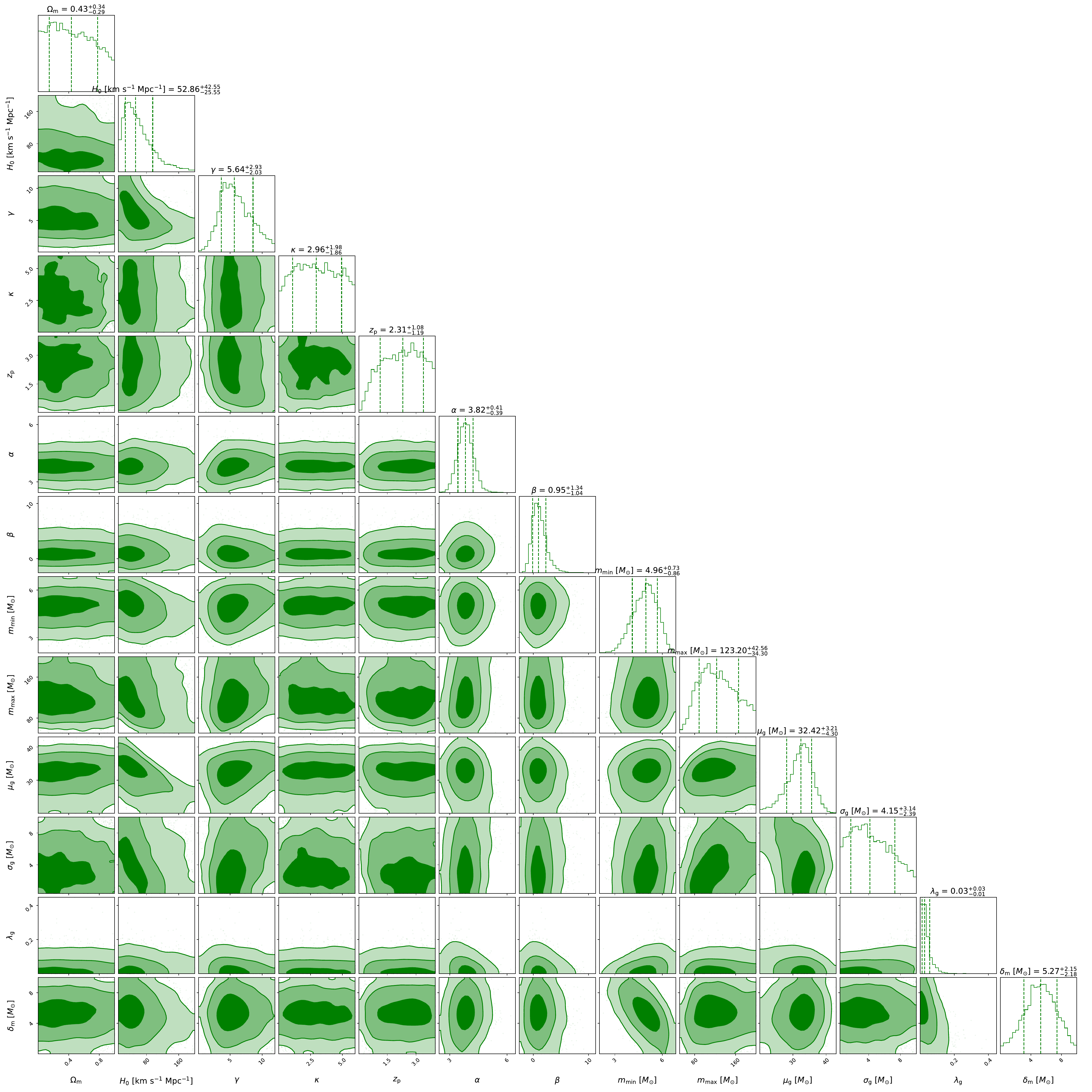}
\caption{Distributions of the parameters for the cosmology model and the \textsc{PowerLaw+Peak} population model (without spin distribution) using the events with SNR$>11$ in GWTC-3. The dashed lines in the marginal distribution represent the median and 68.3\% credible intervals.}
\label{fig:PP_corner}
\end{figure}

We reproduce the results of the \textsc{PowerLaw+Peak} model \citep{2023ApJ...949...76A} in our analysis, to ensure the consistency of our framework with theirs. 
Figure~\ref{fig:PP_corner} shows the corner plots for the posterior distribution obtained using \textsc{PowerLaw+Peak} model with 42 events selected with SNRs $>11$, which is broadly consistent with that of \cite{2023ApJ...949...76A}.
Note that \citet{2023ApJ...949...76A} assumed a Flat$w_0$CDM model with prior of $w_0$ in [-3,0] ({see Table~5 of \citet{2023ApJ...949...76A}}), whereas we assume a Flat$\Lambda$CDM model in this work.

\bibliography{export-bibtex}
\bibliographystyle{aasjournal}

\end{CJK*}
\end{document}